\documentclass[journal]{IEEEtran}

\usepackage{booktabs}
\usepackage{xspace}
\usepackage{microtype}
\usepackage{graphicx}
\usepackage{amsmath,amssymb}
\usepackage{url}
\usepackage[hidelinks]{hyperref}
\graphicspath{{figures/}{paper/figures/}}

\newcommand{\adaptr}{AdaptR4j\xspace}
\newcommand{\rFJ}{Resilience4j\xspace}
\newcommand{\raf}{RAF\xspace}

\begin{document}

\title{When Online Adaptation of Resilience Policies Hurts: Safe Exploration and Measurement Pitfalls in Resilience4j}

\author{Rajat~M.~Barot and Arjun~S.~Borkhatariya%
\thanks{(Corresponding author: Rajat M. Barot.)}%
\thanks{Rajat M. Barot is with the Department of Computer Science, State University of New York,
Binghamton, Binghamton, NY, USA. E-mail: rbarot2@binghamton.edu}%
\thanks{Arjun S. Borkhatariya is with the Department of Software Engineering, Arizona State
University, Tempe, AZ, USA. E-mail: aborkhat@asu.edu}%
\thanks{Code and data: \url{https://github.com/rajatbarot24/adaptr4j}.}}

\markboth{Preprint, 2026}{Barot and Borkhatariya: When Online Adaptation of Resilience Policies Hurts}

\maketitle

\begin{abstract}
Circuit breaker, retry, and bulkhead settings in Java microservices are tuned once and frozen,
although the best setting depends on the fault a dependency is currently showing. We study
whether an online learner can pick these settings at runtime. \adaptr swaps \rFJ policies
atomically in Spring Boot~3 and lets a contextual bandit choose among complete
CircuitBreaker/Retry/Bulkhead configurations from Prometheus signals.

Our first, naive design scored \emph{worse than keeping the default policy} in three of four
winnable fault scenarios under the reward it optimised: exploring a menu that contained aggressive circuit breakers rejected
most traffic while the learner was still learning. Two safeguards fix this: a \emph{safe policy
menu} (no breaker stricter than the latency SLO) and \emph{safe exploration} with a bounded
reward. With them, \adaptr beats the production default in every winnable scenario
($p<0.001$, three runs each, run-to-run SD $\le 0.03$), beats the best fixed policy when the
fault changes over time, and cuts user-visible failures up to 3.7$\times$ compared with the
naive design. A context-free bandit on the same safe menu performs statistically
indistinguishably, so the safeguards---not the learner---carry most of the benefit.

We also report how an earlier version of our harness produced convincing but wrong results:
a unit mismatch made every circuit breaker 100$\times$ too strict and rejected calls were
invisible to the metrics, so 94\,\% of user requests failed while latency and error metrics
looked excellent.
\end{abstract}


\section{Introduction}
\label{sec:intro}

\rFJ~\cite{resilience4j} is the standard fault-tolerance library for Java microservices.
Its CircuitBreaker (CB), Retry, and Bulkhead patterns interact: retries amplify load on a
degraded dependency, a strict breaker turns a brief error burst into a full outage, and a
small bulkhead rejects healthy traffic under load~\cite{sedghpour2022,meiklejohn2022}.
Teams set these parameters once in YAML and freeze them, yet the best setting depends on
whether the dependency is currently slow, failing, or healthy. Adaptive approaches exist at
the service-mesh layer~\cite{sedghpour2023,sun2024} and for single knobs such as concurrency
limits~\cite{netflixlimits,envoyadaptive}, but not for choosing whole \rFJ configurations
inside the application.

We built \adaptr to test whether an online learner can do this. This paper reports where that
attempt stands, including what went wrong:

\begin{itemize}
\item \textbf{Naive online adaptation can be worse than doing nothing.} A LinUCB
  bandit~\cite{li2010} over a 30-item policy menu lost to the production-default policy in three
  of four winnable scenarios, because exploration repeatedly applied breakers that rejected
  almost all traffic (\S\ref{sec:naive}).
\item \textbf{Two simple safeguards make it reliably beneficial.} A configuration-only rule for
  the menu and safe exploration with a bounded reward let \adaptr beat the default everywhere
  it can be beaten and beat the best fixed policy under time-varying faults
  (\S\ref{sec:results}).
\item \textbf{The learner matters less than the safeguards.} A context-free UCB1 bandit on the
  same safe menu is statistically indistinguishable from \adaptr; context features help in one
  scenario and hurt in another (\S\ref{sec:attribution}).
\item \textbf{Measurement can invert conclusions.} Our first harness reported excellent latency
  and error rates while 94\,\% of user requests failed (\S\ref{sec:measurement}). We describe the
  two bugs and the checks that catch them.
\end{itemize}

\section{AdaptR4j}
\label{sec:design}

\textbf{Atomic policy swap.} Spring Boot~3's configuration refresh does not reach \rFJ
registry instances at runtime. A policy holder keeps the active (CB, Retry, Bulkhead) triple
behind an atomic reference; each call decorates the downstream call in \rFJ's standard order
Retry(CB(Bulkhead(call))), and a swap publishes newly built instances with one atomic write.
Calls already in flight finish on the old instances. Retry does not retry the library's own
rejections (open breaker, full bulkhead).

\textbf{Controller.} Every 60\,s the controller reads six Prometheus features (edge p99 and
p50 over the SLO, error rate, retry amplification factor \raf, CB-open flag, normalised
throughput), selects a policy with LinUCB~\cite{li2010}, and pushes it to the service. The
first five windows are shadow mode (observe only). The reward for a window is
\[
  r = \begin{cases}
    -2 & \text{if } p99 > 500\,\text{ms,}\\
    \max(-2,\, u) & \text{otherwise,}
  \end{cases}
\]
\[
  u = 1 - \frac{p99}{500\,\text{ms}} - \frac{\text{err}}{0.05} - 0.5\max(0,\raf-1),
\]
so latency headroom earns up to $+1$, an error rate at the 5\,\% SLO costs a full point, and no
window is worse than an SLO breach. Latency is measured at the edge (the request the user
made), and every call that fails for any reason---including CB and bulkhead rejections---counts
as an error (\S\ref{sec:measurement}).

\textbf{Guardrails.} A RAF hard stop freezes the policy while retries exceed twice the request
rate, and a rollback switches to the best policy seen so far after three windows with
reward $<-1$.

\textbf{Safeguards added after the naive design failed} (\S\ref{sec:naive}).
\emph{Safe menu:} we keep only policies whose CB failure-rate threshold is at least 45\,\%
and whose slow-call duration exceeds the 500\,ms latency SLO. A stricter breaker counts
SLO-compliant calls or ordinary error bursts as failures and opens on healthy traffic. The
rule keeps 14 of the 30 menu policies, including the best fixed policy of every scenario.
\emph{Safe exploration:} a policy whose recent mean reward is more than 1.0 below the best
policy's is not selected again until it has rested 30 windows, so it can be re-tested after
the fault changes. This is a simple form of conservative exploration~\cite{wu2016conservative,garcia2015}.
\emph{Bounded reward:} before, the error term was unbounded and one fully rejecting window
scored about $-20$, dominating every mean.

\section{Experimental Setup}
\label{sec:setup}

\textbf{Testbed.} Service A (edge, hosts \adaptr's swap mechanism) calls Service B, whose
latency and error rate are set at runtime through an HTTP endpoint. Both are Spring Boot
3.3.5 / \rFJ 2.2.0 on Java~21, on one laptop (Intel i5-8250U, 4 cores,
8\,GB RAM, Windows~11). Prometheus 2.52 scrapes every 15\,s. Locust 2.43 drives 50 users
with 0.5--2\,s think time; 200 warm-up requests precede every run. Each run lasts 80 one-minute
windows from a fresh start of both services.

\textbf{Fault scenarios.} S1 moderate (50\,ms, 4\,\% errors); S2 severe (200\,ms, 10\,\%);
S3 pure latency (500\,ms, 0\,\%); S4 error storm (0\,ms, 30\,\%); S5 latency spikes
(0\,$\leftrightarrow$\,500\,ms every 5 minutes). The SLO is p99\,$\le$\,500\,ms and
error rate\,$\le$\,5\,\%. S3 cannot be won: every call already takes 500\,ms, so every policy
misses the latency SLO. We keep it as a sanity check.

\textbf{Baselines.} \emph{Default:} the menu's balanced policy p01 (3 attempts, 500\,ms wait,
CB 50\,\%, 40 concurrent calls), applied for the whole run. \emph{Best fixed policy} (oracle):
a 65-window sweep applies every menu policy for two windows; the best one in hindsight is then
run for 80 windows. \emph{UCB1}~\cite{auer2002} on the safe menu, without context.
\emph{No-context:} \adaptr with a constant context vector. \emph{Naive design:} \adaptr with the
full 30-item menu, no safe exploration, rewards re-scored with the bounded reward.

\textbf{Statistics.} Consecutive windows of a run are autocorrelated, so we compare runs with a
moving-block bootstrap (blocks of 10 windows, 5\,000 resamples) on the difference of mean rewards.
\adaptr was run three times per winnable scenario; baselines once.

\section{How Measurement Fooled Us}
\label{sec:measurement}

The first version of this study reported mean rewards of $+0.72$ (S1) and $+0.81$ (S2), p99
reductions of 9--12$\times$ over the default, and error rates of 0.1\,\%. Comparing the
controller's view with the load generator's view revealed that \textbf{94.5\,\% (S1) and
97.4\,\% (S2) of user requests had failed}, with HTTP~500 (Table~\ref{tab:measurement}).
Two bugs combined:

\textbf{1. Unit mismatch.} The policy menu expressed CB thresholds as fractions (0.50); \rFJ
expects percentages. Every policy therefore ran with a 0.2--0.8\,\% threshold, and each breaker
opened on the first injected error. The only configuration that served users well was an
admission-control baseline that happened to send its threshold as \texttt{90.0}.

\textbf{2. Invisible rejections.} Latency and errors were recorded on the downstream call
inside the resilience chain. Calls rejected by an open breaker never reached it: they added
near-zero durations, which pulled p99 down, and were never counted as errors. The reward
therefore preferred the policies that rejected the most traffic.

\begin{table}[t]
\centering
\small
\caption{What the controller saw vs.\ what users experienced (\adaptr; per-window p99 and error
  rate averaged over the run; users failed = Locust, whole run).}
\label{tab:measurement}
\begin{tabular}{llrrr}
\toprule
Harness & Scen. & p99 seen & errors seen & users failed \\
\midrule
original & S1 & 138\,ms & 0.10\,\% & \textbf{94.5\,\%} \\
original & S2 & 106\,ms & 0.06\,\% & \textbf{97.4\,\%} \\
fixed    & S1 & 280\,ms & 0.12\,\% & 0.1\,\% \\
fixed    & S2 & 854\,ms & 3.64\,\% & 6.0\,\% \\
\bottomrule
\end{tabular}
\end{table}

Three smaller defects made it worse: a non-standard decorator order (the bulkhead held its slot
through every retry back-off), a load-generator shape class that silently overrode the requested
user count, and a rollback that counted shadow-mode windows. After fixing all of them, the
controller's error rate and the users' failure rate differ by about 0.1 percentage points on
average in validation runs.

\textbf{Checks we now apply.} (1) Measure at the outermost layer the user sees; count every
exception, including the library's own rejections. (2) Cross-check controller metrics against
the load generator in every run, automatically. (3) Read back the live configuration from the
library (\texttt{/actuator/circuitbreakers}) after every policy change. (4) Treat results that
look \emph{too} good---near-zero p99 under injected latency---as a bug until explained.

\section{Results}
\label{sec:results}

\subsection{The naive design loses to the default}
\label{sec:naive}

On the corrected harness, the naive design (full 30-item menu) scores $+0.44$ in S1 but
$-1.86$, $-0.85$, and $-0.76$ in S2, S4, and S5 (Table~\ref{tab:main}). With the original
unbounded reward, it was worse than the default p01 in S2, S4, and S5. The sweep explains why:
16 of the 30 policies let more than half of all requests fail in at least one scenario, all of
them because of strict breakers, and exploration kept returning to them.

\begin{table*}[t]
\centering
\small
\caption{Mean reward (higher is better; $-2$ = SLO missed every window) and user-visible failure
  rate (Locust: HTTP~5xx or response $>1$\,s). \adaptr: three runs, mean\,$\pm$\,SD; ``late'' =
  windows 40--80. $^\dagger$: block-bootstrap $p<0.05$ vs.\ \adaptr run~1.}
\label{tab:main}
\begin{tabular}{l rrr rr rr rr rr}
\toprule
 & \multicolumn{3}{c}{\adaptr (safe design)} & \multicolumn{2}{c}{Default p01}
 & \multicolumn{2}{c}{Best fixed policy} & \multicolumn{2}{c}{UCB1, safe menu}
 & \multicolumn{2}{c}{Naive design} \\
\cmidrule(lr){2-4}\cmidrule(lr){5-6}\cmidrule(lr){7-8}\cmidrule(lr){9-10}\cmidrule(lr){11-12}
Scen. & reward & late & fail\% & reward & fail\% & reward & fail\% & reward & fail\% & reward & fail\% \\
\midrule
S1 & $+0.257 \pm 0.002$ & $+0.518$ & 0.1 & $-2.000^\dagger$ & 0.1 & $+0.518$ & 0.0 & $+0.050$ & 0.8 & $+0.441$ & 0.3 \\
S2 & $-1.821 \pm 0.021$ & $-1.759$ & 5.8 & $-2.000^\dagger$ & 1.0 & $-1.438^\dagger$ & 9.4 & $-1.823$ & 4.0 & $-1.862$ & 9.4 \\
S3 & $-2.000$ (1 run)   & $-2.000$ & 0.8 & $-2.000$ & 0.0 & $-2.000$ & 0.0 & -- & -- & $-2.000$ & 10.6 \\
S4 & $-0.315 \pm 0.010$ & $-0.035$ & 9.4 & $-2.000^\dagger$ & 32.5 & $+0.249^\dagger$ & 2.6 & $-0.526$ & 14.0 & $-0.848$ & 31.4 \\
S5 & $-0.451 \pm 0.026$ & $-0.409$ & 2.3 & $-0.842^\dagger$ & 0.0 & $-0.763^\dagger$ & 0.0 & $-0.603$ & 0.4 & $-0.764$ & 8.5 \\
\bottomrule
\end{tabular}
\end{table*}

\begin{figure*}[t]
\centering
\includegraphics[width=0.8\textwidth]{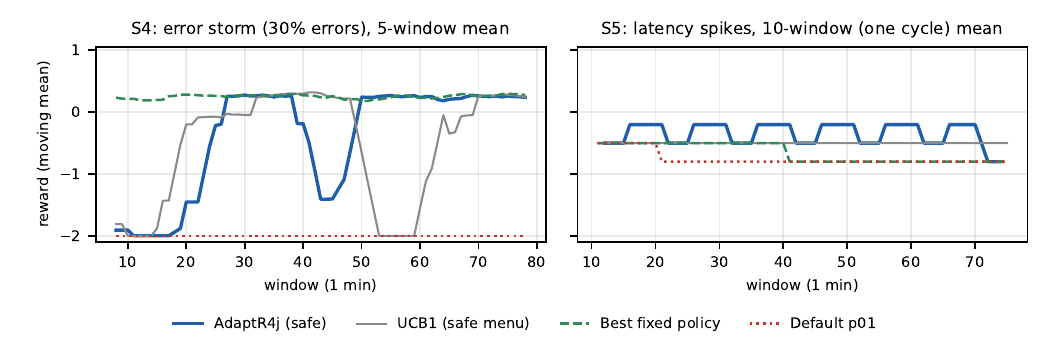}
\caption{Reward over time (moving mean) in S4 and S5, run~1. In S4 \adaptr reaches the best fixed
policy's level after about 25 windows; in S5, where the fault alternates, it stays above every
fixed policy. Default p01 stays at $-2$ throughout S4.}
\label{fig:timeline}
\end{figure*}

\subsection{The safe design}

\textbf{Reproducible.} Across three independent runs the mean reward has a standard deviation of
at most 0.026, and every run's most frequent policy is the same (p24 in S1 and S4, p04 in S2,
p21 in S5). In S1 and S4 it settles on p24 (84\,\% and 74\,\% of windows); in S2 and S5 it keeps
switching, as no single policy is clearly best.

\textbf{Better than the default everywhere it can be.} \adaptr beats p01 in S1, S2, S4, and S5
($p<0.001$ each). The default misses the latency SLO in every window of S1, S2, and S4 because
its retry waits push p99 to 0.7--1.1\,s. It does have fewer user-visible failures in S2 and S5:
it trades latency for success, while the reward weighs both (\S\ref{sec:open}).

\textbf{Beats the best fixed policy when the fault changes.} In S5, latency alternates every five
minutes, and no fixed policy suits both phases. \adaptr scores $-0.451$ against $-0.763$ for the
best fixed policy ($p=0.011$) and $-0.842$ for the default (Figure~\ref{fig:timeline}, right).

\textbf{Approaches the best fixed policy under steady faults, slowly.} In S1 \adaptr settles on
the best fixed policy p24 and matches it exactly after window~40 ($+0.518$), but trying all 14
policies first costs reward: its overall mean ($+0.257$) is below the naive design's ($+0.441$),
which happened to try only five policies before settling on a good one (window~10 vs.\ 19). In S4 it reaches the best fixed policy's
level after about 25 windows (Figure~\ref{fig:timeline}, left); the dip at windows 41--47 is the
scheduled re-test of policies excluded about 30 windows earlier. In S2 the best fixed policy remains clearly better
($-1.44$ vs.\ $-1.82$).

\textbf{Fewer user-visible failures than the naive design}: 9.4\,\% vs.\ 31.4\,\% in S4, 2.3\,\%
vs.\ 8.5\,\% in S5, and 5.8\,\% vs.\ 9.4\,\% in S2.

\subsection{Safeguards vs.\ learner}
\label{sec:attribution}

UCB1 on the same safe menu, without any context, scores $+0.050$, $-1.823$, $-0.526$, and
$-0.603$ in S1, S2, S4, and S5. \adaptr's means are higher by 0.21, 0.00, 0.21, and 0.15, but none
of these differences is significant under the block bootstrap ($p=0.34$, $0.79$, $0.47$, $0.17$):
reward varies strongly from window to window, and each baseline was run once. The safe menu, by
contrast, helps every learner: UCB1 improves from $-0.353$ to $+0.050$ in S1 when moving from the
full to the safe menu. Context features are not clearly useful either: removing them lowers the
mean in S1 ($+0.186$ vs.\ $+0.257$) but raises it in S4 ($-0.140$ vs.\ $-0.315$).
Our evidence therefore supports the safeguards, not contextual learning, as the source of
the improvement.

\section{Related Work}

Sedghpour et al.\ show that no static circuit-breaker setting dominates across
workloads~\cite{sedghpour2022} and adapt service-mesh breakers with rules~\cite{sedghpour2023};
Sun et al.\ use deep Q-learning in Istio~\cite{sun2024}. RetryGuard~\cite{retryguard2025}
prevents retry storms with an analytical model. Netflix's concurrency limits~\cite{netflixlimits},
Envoy's adaptive concurrency filter~\cite{envoyadaptive}, client-side
throttling~\cite{sre2016}, and admission control such as TopFull~\cite{topfull2024} and
Rajomon~\cite{xing2025} adapt a single knob. Autothrottle~\cite{autothrottle2024} and
OPPerTune~\cite{oppertune2024} learn resource and configuration settings online. \adaptr instead
chooses whole \rFJ configurations inside the application, as a MAPE-K
loop~\cite{kephart2003,garlan2004}. Our negative first result echoes the safe-RL
literature~\cite{garcia2015,wu2016conservative}: unconstrained exploration on a live system can
cost more than adaptation recovers.

\section{Open Issues and Next Steps}
\label{sec:open}

\textbf{The menu rule was found after the fact.} We derived the safe-menu rule from the same
sweeps we evaluate on. It uses only configuration values an operator knows in advance, but it
must be validated on new services and fault types.

\textbf{Exploration cost.} Trying every safe policy once costs 15--20 windows; S1 shows the price.
Seeding the bandit with the default's observed reward, or with offline sweeps, is the obvious
next step.

\textbf{What should the reward value?} In S2 and S5 the default serves more requests
successfully but slowly; our reward prefers fast partial failure. The right trade-off is a
service-level decision and should be configurable.

\textbf{No feasible policy in S2 and S3.} When no menu policy can meet the SLO, the controller can
only choose the least bad one; menu generation should cover such regimes.

\textbf{Threats to validity.} All runs are on one laptop with synthetic faults and one workload;
baselines were run once; the fan-out, drift, and network-latency experiments we ran with the naive
design have not yet been repeated with the safe design.

\textbf{Reproducibility \& availability.} Code, policy menus, all decision logs and load-generator
outputs, and the scripts that produce every number in this paper are available at
\url{https://github.com/rajatbarot24/adaptr4j} under the MIT license.

\section{Conclusion}

Online selection of \rFJ policies is only beneficial once exploration is made safe. With a safe
policy menu, safe exploration, and a bounded reward, a simple bandit beats the production default in
every winnable scenario we tested and beats the best fixed policy when faults change over time;
without them, it is worse than doing nothing. Equally important, our own first evaluation shows how
easily resilience experiments can report excellent metrics while most users fail.

\bibliographystyle{IEEEtran}
\bibliography{refs}

\end{document}